\documentclass[doublecol]{epl2}

\usepackage{amsmath}
\usepackage{epstopdf}
\usepackage{color}

\usepackage{epsfig}
\usepackage{graphicx}
 
\usepackage{amssymb}
\usepackage{amsfonts}
\usepackage{dcolumn}

\title{Dissipative surface solitons in periodic structures}

\author{Yaroslav V.  Kartashov\inst{1}, Vladimir V. Konotop\inst{2}, \and Victor A. Vysloukh\inst{3}}
\shortauthor{Y. V. Kartashov, V. V. Konotop, and V. A. Vysloukh}

\institute{\inst{1}ICFO-Institut de Ciencies Fotoniques, and Universitat Politecnica de Catalunya, Mediterranean Technology Park, 08860 Castelldefels (Barcelona), Spain
 \\
\inst{2}Centro de F\'{\i}sica Te\'orica e Computacional
  and Departamento de F\'{\i}sica, Faculdade de Ci\^encias, Universidade de Lisboa,
Avenida Professor Gama
Pinto 2, Lisboa 1649-003, Portugal
\\
$^3$Departamento de Fisica y Matematicas, Universidad de las Americas - Puebla, Santa Catarina Martir, 72820, Puebla, Mexico
}

\pacs{42.65.Tg}{Optical solitons; nonlinear guided waves }
\pacs{42.65.Jx}{Beam trapping, self-focusing and defocusing; self-phase modulation }
\pacs{42.65.Wi}{Nonlinear waveguides }

\abstract{  
We report dissipative surface solitons forming at the interface between a semi-infinite lattice and a homogeneous Kerr medium. The solitons exist due to   balance between amplification in the near-surface lattice channel and  two-photon absorption. The stable dissipative surface solitons exist in both focusing and defocusing media, when propagation constants of corresponding states fall into a total semi-infinite and or into one of total finite gaps of the spectrum (i.e. in a domain where propagation of linear waves is inhibited for the both media). In a general situation, the surface solitons form when amplification coefficient exceeds threshold value. When a soliton is formed in a total finite gap there exists also the upper limit for the linear gain.  
}

\begin{document}

\maketitle

%\noindent
Shallow periodic modulations of a refractive index  may considerably affect the diffraction of low-power beams and properties of stationary nonlinear excitations~\cite{r1}. Of particular interest are truncated periodic structures that support so-called surface lattice solitons propagating along the interfaces between periodic and uniform media. While the history of linear surface modes goes back to the seminal paper of Tamm~\cite{Tamm},  one-dimensional  surface modes in nonlinear lattices were predicted in~\cite{r3} and subsequently obtained in both focusing \cite{r4} and defocusing \cite{r5,r6} materials, while their two-dimensional counterparts were generated in \cite{r8}. A characteristic feature of surface solitons is that they usually exist above the energy flow threshold. Surface lattice solitons inherit some properties of solitons in bulk lattices, in a sense that their shapes and transverse extent depend on the location of propagation constant in a band-gap lattice spectrum. Moreover, there exists no sharp transition between  modes strongly localized at the surface and modes localized at some distance from the surface or bulk modes: the localization domain moves smoothly from the boundary to some domain in the bulk as one follows the one-parametric branch of the solutions ~\cite{BludKon}. A surface can also support breather solutions, i.e. localized modes with periodically changing  shapes~\cite{BludKon}. Even more rich behavior is observed when a defect is added  into otherwise periodic guiding structure. In particular, when a defect is introduced into a surface lattice channel it may dramatically affect the conditions for surface soliton formation and result in appearance of new types of surface modes~\cite{r16}.

The above referred studies were dedicated to conservative systems. 
An interesting and still open problem is the impact of non-conservative surfaces  on existence and properties of the localized modes. A particularly interesting situation arises when a gain is applied at the interface between a homogeneous and a periodic medium. It is relevant to mention that the effect of a localized gain was analyzed for existence and interaction of gap solitons in shallow fiber Bragg gratings \cite{r20},
% and in periodic media with nonlinear losses~\cite{KKVT},  
in media of two-level atoms \cite{r21}, for formation of spatial dissipative solitons in systems governed by complex Ginzburg-Landau equation \cite{r22}, and for dynamic emission of moving lattice solitons in systems without dissipation \cite{r23}. 

In this Communication we address the properties of truly stationary dissipative solitons forming at the edge of semi-infinite lattice and existing due to the balance between localized gain in the near-surface lattice channel and strong two-photon absorption in a cubic medium. Dissipative surface solitons~\cite{r24} were studied before only in the truncated discrete systems governed by Ginzburg-Landau equation with uniform gain (the  model  introduced in~\cite{r26}). Here we show that in experimentally realistic setting dissipative surface solitons are pinned by the "defect" channel where gain is realized. Such modes are attractors and thus can be excited from sufficiently large class of initial conditions, ranging from localized or extended regular patterns to noisy patterns. We argue that  surface solitons form when gain coefficient exceeds a threshold value, and if the propagation constant belongs to a finite gap, there exist also the upper bound  for the allowed gain coefficient at which surface solitons can still be found.

Specifically, we consider propagation of laser radiation in a
lattice with spatially localized linear gain and nonlinear losses
that can be described by the nonlinear
Schr\"odinger (NLS) equation for the dimensionless light field amplitude
$q$:
\begin{eqnarray}
iq_\xi = - \frac 12 q_{\eta\eta} -\left[ R(\eta) -i \gamma(\eta)\right]q-\sigma |q|^2q - i\alpha |q|^2q.
\label{GPEq}
\end{eqnarray}
 {Here $\eta=x/x_0$ and $\xi=z/L_{dif}$ are the   transverse and longitudinal coordinates normalized  to the characteristic beam    width $x_0$ and to the diffraction length $L_{dif}=kx_0^2$,  $k=2\pi n_0/\lambda$, $\lambda$ is the wavenumber, $n_0$ is the unperturbed refractive index,  $I_0$ is the characteristic intensity, $L_{nl}=n_0/kn_2I_0$   is the nonlinear self-action length, $p_i=L_{dif}/L_{gain}$is linear gain coefficient, $L_{gain}
%=1/\alpha_1
$ is the amplification length, $\alpha=L_{dif}/L_{loss}$ is the coefficient of nonlinear losses,    $L_{loss}=1/\alpha_2I_0$ characterizes the length of two-photon absorption and  $\sigma=1$ ($\sigma=-1$) corresponds to focusing (defocusing) nonlinearity.
}

 {
Spatial solitons were successfully observed in nonlinear waveguide arrays made of AlGaAs alloy below the half-band-gap, $\lambda\approx 1.53$ $\mu$m.
Such arrays have typical length of 6 mm, waveguide separation of $4\div 7$ $\mu$m  and effective core area about 
$20$ $\mu$m$^2$~\cite{Exper}. In this case (as well as in our model) the two-photon absorption is the dominating mechanism of optical losses: the linear absorption coefficient is around 0.1 cm$^{-1}$ while the typical value of two-photon absorption coefficient is $\alpha_2=0.3$ cm/GW, and the representative peak soliton intensity is above $I_0\approx 5$ GW/cm$^2$.
Importantly, the same semiconductor material (AlGaAs) is widely used for production of the wide-band (typical bandwidth of $60\div 70$ nm) semiconductor optical amplifiers  with rather high optical gain 
%(around few inverse millimeters) 
in the same spectral range~\cite{Exper2}.
In particular,  for the above mentioned structure,  $n_0=3.34$ and the Kerr coefficient $n_2=1.6\cdot 10^{-13}$ cm$^2$/W. Then for a laser beam with $x_0=5$ $\mu$m  one obtains  $L_{dif}\approx L_{nl}\approx 0.34$ mm and 
%the length of two-photon absorption 
$L_{loss}\approx 6.67$ mm, that corresponds to $\alpha\approx 0.05$ (TE polarization is under consideration). Then $p_i=0.1$  is achieved for $L_{gain}\approx 0.34$ cm.  
}

While qualitatively the results reported below are valid for general periodic modulations $R(\eta)$,  the quantitative analysis is performed for the semi-infinite lattice of the form  $R(\eta)= p_r\sin^2(2\eta)$, with $p_r$ being the depth of the lattice proportional to the refractive index modulation,  which is placed at  $\eta>0$ (notice that the period of the structure can be also made $\pi/2$ by simple renormalization). At $\eta \leq 0$ we have a homogeneous medium where $R(\eta)\equiv 0$.  We suppose that localized gain, whose profile is described by the function $\gamma(\eta)$,  is realized in the vicinity of a near-surface lattice channel. In the simplest case we consider $\gamma(\eta)$ exactly coinciding with $R(\eta)$ in the first channel of the lattice, i.e. $\gamma(\eta)= p_i\sin^2(2\eta)$ 
($p_i>0$ is the linear gain coefficient)   for  $0<\eta<\pi/2$, and $\gamma(\eta)\equiv 0$ otherwise, but the situation when the peak position of $\gamma(\eta)$ is shifted with respect to first maximum of $R(\eta)$ by a distance $\eta_s$  will be considered too. 
%It should be pointed out that bulk solitons in more sophisticated periodic and parabolic gain landscapes were %considered in \cite{r28,r29}.

Localized dissipative solitons are searched in the form $q(\eta,\xi)=w(\eta)\exp[ib\xi+i\theta(\eta)]$  with $w(\eta)$ and $\theta(\eta)$ being the amplitude and the real stationary phase of the field, and $b$  being the propagation constant.  Exponentially localized modes in periodic media~\cite{AKS} and conservative surface modes~\cite{r3,r4,r5}
%, and dissipative defect modes in lattices~\cite{KKVT}  
emerge when the propagation constant 
$b$ 
falls into one of the gaps of spectrum of periodic guiding structure. It turns out that this is also true in the case of dissipative surface solitons. 
To prove this, we rewrite (\ref{GPEq}) in terms of the real functions
$w(\eta)$ and $v(\eta)\equiv\theta_\eta$:
\begin{subequations}
\label{rho_theta}
\begin{eqnarray}
\label{rho}
&&  w_{\eta\eta} -2bw- v^2 w 
% w\theta_\eta^2
+2\sigma w^3+2Rw=0, 
%\end{eqnarray}
%\begin{eqnarray}
%\label{theta}
\\
\label{theta}
&&
(vw^2)_\eta
%\theta_\etaw^2
-2\gamma w^2+2\alpha w^4=0
\end{eqnarray}
\end{subequations}
and consider first the limit $\eta\to-\infty$. Since in this limit $R\equiv 0$,  one readily finds the explicit asymptotics 
\begin{subequations}
\label{asymp}
\begin{eqnarray}
\label{asymp_w}
&& w=A_-e^{\sqrt{2 b}\eta}- \sigma A_-^3e^{3\sqrt{2 b}\eta}/(8b)+{\cal O}(e^{5\sqrt{2 b}\eta})
\\
\label{asymp_v}
&& v=-\alpha A_-^2e^{2\sqrt{2 b}\eta}/(2\sqrt{2 b})+{\cal O}(e^{4\sqrt{2 b}\eta})
\end{eqnarray}
\end{subequations}
where  $A_-$ is a real constant, depending on the total energy flow $U=\int_{-\infty}^{\infty}w^2d\eta$.
The first (trivial) consequence of the obtained asymptotics is that localized solutions exist only in domains outside the linear spectrum, i.e. at $b>0$. Second,  important conclusion   is that in the asymptotic region  $ v^2w\sim \exp(5\sqrt{2 b}\eta) $, i.e. decays faster than $w^3$. Thus, at $\eta\to-\infty$  the shape of the mode is described by the conservative properties of the medium, i.e. $q(\eta,\xi)$ exponentially approaches the standard stationary NLS soliton.  

\begin{figure}[h]
  %\begin{center}
%\includegraphics[width=0.5\textwidth]{fig1.eps}
\epsfig{file=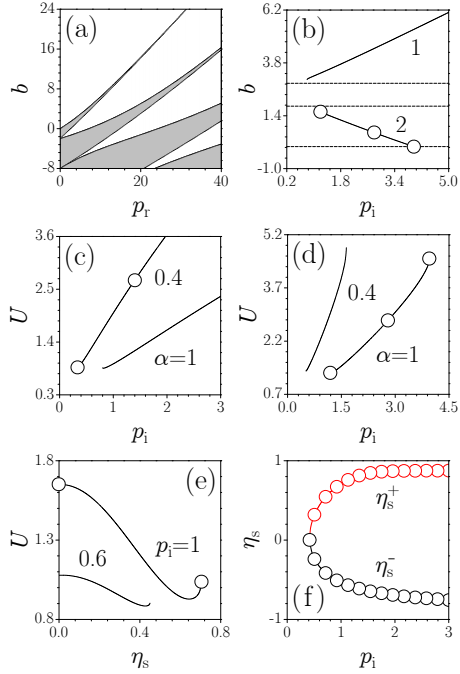,width=\columnwidth}
  %\end{center}
  \vspace{-0.5cm}
\caption{(a) Band-gap spectrum of the infinite  lattice.   (b) $b$ {\it vs} $p_i$ in focusing (curve 1) and defocusing (curve 2) media at $\alpha=1$. The horizontal lines show the limits of the total gaps. $U$ {\it vs} $p_i$ in focusing (c) and defocusing (d) media.  Panels (b)-(d) correspond to $\eta_s=0$. (e) $U$ {\em vs} $\eta_s$  at $\alpha=0.5$.  (f) Maximal positive ($\eta_s^+$)  and negative ($\eta_s^-$)  shifts of amplifying domain with respect to $R(\eta)$ at which surface soliton still exists {\it vs} $p_i$ at   $\alpha=0.5$. The circles in (b) and (e) correspond to solitons shown in Fig.~\ref{fig2}(a) and (b), the circles in (c),(d) correspond to solitons shown in the right column of Fig.~\ref{fig2}.} 
\label{fig1}
\end{figure}

Similar, but more sophisticated, analysis can be performed for the limit $\eta\to\infty$, where the linear lattice is present. One still can prove that $v^2w$ decays faster than  $w^3$, and thus the leading order for the field amplitude is given by the Floquet theorem: $w=A_+\exp(-\mu \eta)P_n(\eta)$ where $P_n(\eta)$ is an  $\pi/2$ or $\pi$ periodic function, $ \mu$ is the Floquet exponent which is determined by the detuning of the propagation constant from the band-edge towards the $n$-th stop gap (the lower and upper boundary of $n$-th gap will be designated by $b_n^{-}$ and $b_n^{+}$, respectively, the number of the first finite gap is set to be $n=1$, while the semi-infinite gap is denoted by $n=0$), and $A_+$ is the  normalization constant (see e.g.~\cite{AKS} for more details). We illustrate the band-gap spectrum in Fig.~\ref{fig1}(a).

Thus, for existence of a localized surface mode one has to require that the propagation constant lies 
beyond the allowed band of the linear spectrum of the uniform medium, on the one hand,   and belongs to one of the gaps of the periodic structure on the other hand. The respective domains will be termed total gaps (to distinguish them form the own gaps of the lattice: as it is clear total gaps represent a subset of the lattice gaps). For example, for the case of $p_r=5$, studied below in details, there are only two total gaps:
 the semi-infinite total gap $b\in (2.875,\infty)$ and the total finite gap $(0,1.840)$ while the  lattice gaps, shown in Fig.~\ref{fig1}(a), are  given by $(2.875,\infty)$,  $(-0.645,1.840)$, etc. (i.e. $b_0^{-}=2.875$, $b_1^{+}=1.840$, $b_1^{-}=-0.645$ etc.)
  
%This situation is illustrated in Fig.~\ref{fig0}. 
%\begin{figure}[h]
  %\begin{center}
%\includegraphics[width=0.5\textwidth]{fig1.eps}
% \vspace{-0.5cm}
%\epsfig{file=fig0,width=\columnwidth}
%  %\end{center}
%  \vspace{-1.5cm}
%\caption{Schematic presentation of the band-gap structure for the lattice $\eta>0$, computed for $p_r=5$ and  %the homogeneous medium at $\eta<0$. Shadowed domains indicate allowed bands. For a localized mode $b$ must %belong to a total gap, i.e. to one of the empty domains  outlined by one solid and one dashed  horizontal %lines and by bold vertical lines.  For $p_r=5$  there is a total semi-infinite gap and one total finite gap.} 
%\label{fig0}
%\end{figure}

The established constrains on  $b$, naturally impose limitations on possibility of the excitation of the surface modes. The mismatch between the boundaries of the gaps in the left and right hand structures implies the existence of a threshold value  $U_{cut}$, of the energy flow. In order to show this, let us consider a mode whose propagation constant $b$ belongs to the total semi-infinite gap  and approaches $b_0^{-}$, i.e. $0<b- b_0^{-}\ll 1$. Considering $\eta>0$, we observe that when $b\to b_0^{-}$ the amplitude $w$ tends to zero: $ w_{\max}\to 0$ (see e.g.~\cite{PhysicaD} and references therein). On the other hand,  considering (\ref{rho_theta}) at $\eta<0$ as an ODE defining the shape of the soliton, the smallness of $w$ means that the terms $v^2w$ in (\ref{rho}) can be neglected and the asymptotic behavior is described by the conservative NLS equation, i.e. by (\ref{asymp_w}). Since,  $b$ does not go to zero (due to finite value of $b_0^{-}$), also the nonlinear term $w^3$ can  be neglected and the field behavior is described by the linear equation $w_{\eta\eta}=2bw$. Thus the function $w$ must be exponentially decaying, what for the linear ODE at hand is only possible if $w_\eta(0)=\sqrt{2b}w(0)$. As it is clear this is an extra condition, in addition to the continuity of $q$ and $q_\eta$, which must be satisfied at the boundary, i.e. at $\eta=0$. In a general situation this is impossible with only two available constants $A_+$ and $A_-$, which are determined by the total energy $U$   and by the properties of the linear lattice at $\eta>0$.   In other words, by assuming that the amplitude of the mode can go to zero we have arrived at a contradiction. Thus, there exists a minimal threshold value of $w_{\max}$ above which surface modes can exist. Taking now into account the Sobolev inequality $ | w_{\max}|^2\leq 2 U\int |w_\eta|^2d\eta$ we  conclude that there exists a threshold for the energy flow. Moreover, since the limit $b\to b_0^{-}$ would imply $w_{\max} \to 0$ we finally conclude that there exits also a cut-off value $b_{cut}>b_0^{-}$ such that only for $b>b_{cut}$ one can find dissipative surface modes. Below we illustrate these properties in numerical examples. We notice that the above arguments are also valid for the pure conservative case, and thus explain the threshold values for the energy flow $U$, observed in earlier studies~\cite{r3,r4,r5}.

\begin{figure}[h]
  \begin{center}
\epsfig{file=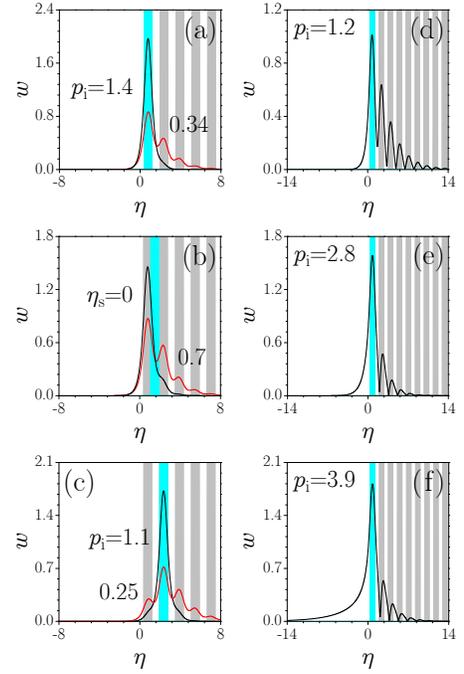,width=\columnwidth}
% \vspace{0.5cm}
   \end{center}
\caption{Surface solitons in focusing (left column) and defocusing (right column) media. Panel (a) shows solitons residing in first lattice channel at $\alpha=0.4$, $\eta_s=0$.  Panel (b) shows solitons residing in first lattice channel for various shifts of amplifying domain with respect to conservative lattice at $p_i=1$,$\alpha=0.5$. Panel (c) shows solitons residing in second lattice channel at $\alpha=0.4$, $\eta_s=\pi/2$. Gray regions indicate guiding lattice channels, while cyan regions show amplifying domains. In (b) we show only amplifying domain corresponding to $\eta_s=0.7$.
%Profiles of gap surface solitons in defocusing medium at $\alpha=1$. Gray regions indicate guiding lattice %channels, while cyan regions show amplifying domains.
}
\label{fig2}
\end{figure}

Dissipative surface solitons exist not only due to balance between diffraction, refraction, and nonlinearity, but also due to  balance between localized gain and nonlinear losses, expressed by the condition 
%\begin{eqnarray}
$
\alpha\int_{-\infty}^{\infty} w^4d\eta=\int_{-\infty}^{\infty}\gamma(\eta)w^2d\eta.
$
%\end{eqnarray}
Hence the propagation constant $b$   and the energy flow $U$  are determined by the gain $p_i$ and by nonlinear losses $\alpha$. Typical dependencies $b(p_i)$ for surface solitons in focusing medium are shown in Fig.~\ref{fig1}(b).  The propagation constant $b$ in the focusing medium falls into the total semi-infinite gap, i.e.  $b>b_0^-$,  and monotonically grows with $p_i$. Such surface solitons exist above some threshold value of $b$, $b_{cut}>b_0^-$,  respectively above the minimal value of the gain coefficient, denoted below as $p_i^{low}$, and above the threshold energy flow $U_{cut}$. The energy flow monotonically increases with $p_i$ everywhere except for a very narrow region close to $p_i^{low}$ [Fig.~\ref{fig1}(c)].   
Typical profiles of dissipative surface solitons in a focusing medium are shown in Fig.~\ref{fig2}(a) when gain is realized in near-surface lattice channel. 
For low values of $p_i$  the  surface solitons expand considerably into the lattice region and acquire shape reminiscent to oscillations of shape of Bloch state bordering the respective gap edge.  
Due to energy flow in the transverse direction and despite the presence of nonlinear losses in the entire medium solitons may extend far beyond the amplifying region. 
With increase of gain $p_i$ the light gradually concentrates in the near-surface channel. It should be stressed that solitons may form in near-surface lattice channel even when the gain  is displaced by a distance $\eta_s$   with respect to the first maximum of the lattice. Typical profile of a surface soliton supported by such shifted gain landscape is shown in Fig.~\ref{fig2}(b) - here the maximum of the field remains in the surface channel despite the fact that gain is realized almost between first and second channels. The energy flow of surface solitons first decreases with $\eta_s$, and then increases when shift approaches maximal value beyond which surface soliton can not form in the first lattice channel [Fig.~\ref{fig1}(e)]. The maximal possible shift of gain landscape quickly increases with $p_i$ and saturates already at $p_i=3$ [Fig.~\ref{fig1}(f)]. Remarkably, surface solitons may form not only when gain profile is shifted into the depth of the lattice (positive $\eta_s$), but also when gain is shifted into uniform medium (negative $\eta_s$). When shift $\eta_s$ becomes sufficiently large solitons may form in second, third, etc, channels of the lattice. Representative examples of profiles of dissipative surface solitons in second lattice channel are shown in Fig.~\ref{fig2}(c). Such solitons feature smaller thresholds (both in terms of $p_i$ and $U$) for their existence than their counterparts in the first lattice channel.

\begin{figure}[h]
\vspace{-1.8cm}
 % \begin{center}
  %\begin{tabular}{c}
%  \includegraphics[width=0.45\textwidth]{fig3_new.eps}
\epsfig{file=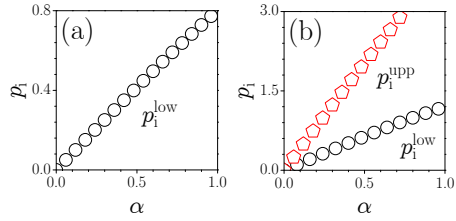,width=\columnwidth}
\vspace{-3.5cm}
   %\end{tabular}
%   \end{center}
\caption{Domains of existence of surface solitons in focusing (a) and defocusing (b) media. In the focusing medium surface solitons exist for all $p_i$  exceeding lower value $p_i^{low}$, while in defocusing medium surface solitons exist between lower $p_i^{low}$ and upper $p_i^{upp}$ values of gain coefficient.}
\label{fig3}
\end{figure}

In the case of defocusing medium localized gain can support dissipative gap solitons featuring characteristic oscillating tails (inside the lattice) at the surface of semi-infinite lattice. The propagation constant of such solitons falls into the finite total gap, $b\in(0, b_1^{+})$ [see Fig.~\ref{fig1} (b) where there is only one total gap] and decreases with $p_i$. Like their counterparts in focusing medium, now 
the solitons emerging from the finite total gap exist
 above the minimal value of gain coefficient $p_i^{low}$ (for this value of linear gain the propagation constant approaches a cut-off value that is close to  the upper edge of the total finite gap). However now, due to finiteness of the gap, there exists also the upper limit for the linear gain $p_i^{upp}$ at which $b$ reaches zero value, i.e. dissipative gap solitons can be found for  $ p_i^{low}\leq p\leq p_i^{upp}$. Respectively, the energy flow takes on the values from the finite interval, where it is the increasing function of $p_i$ [Fig.~\ref{fig1}(d)]. When $p_i\to p_i^{low}$   gap surface solitons expand dramatically into lattice region, but remain well localized inside uniform medium [Fig.~\ref{fig3}(a)], in accordance with the asmyptotcs (\ref{asymp}). The best overall localization is achieved for intermediate $p_i$ values [Fig.~\ref{fig2}(e)].   When $b$ approaches zero (respectively $p_i$ approaches $p_i^{upp}$) the gap surface solitons again become poorly localized due to appearance of long tails in the uniform medium [Fig.~\ref{fig2}(f)].   The domains of existence of dissipative surface solitons in both focusing and defocusing media on the $(\alpha,p_i)$-plane    are shown in Fig.~\ref{fig3}. Minimal gain $ p_i^{low}$ required for the existence of solitons in uniform medium, as well as the width $p_i^{upp}- p_i^{low}$ of the band of gain coefficients where solitons exist in defocusing medium, increase with nonlinear losses $\alpha$.
%

 %
% Fig 03
Finally, we analyzed stability of obtained soliton solutions. We have performed both, the linear stability analysis and direct propagation method in the presence of input perturbations. Both of them showed that the  dissipative surface solitons  are exceptionally robust and can withstand even strong shape deformations almost in the entire existence domain (we considered the lowest branches only).

To conclude, we have demonstrated that an interface with a gain between a periodic and a homogeneous medium can support a diversity of the surface solitons. The propagation constants of the modes belong to one of the total gaps of the structure. Such modes are characterized by the presence of threshold values of the energy flow and cut-off values of the prorogation constant and respectively of the linear gain. The dissipative surface solitons are attractors and therefore can be easily excited from a wide range of initial conditions. Our analysis clearly indicates that similar modes can be obtained in a more general situation where an interface with gain separates two different periodic media. In this last case one can expect larger diversity of the surface modes,e especially when the composite structure is characterized by more that one total finite gap. Finally the reported surface dissipative solitons seem to be very promising objects for spectroscopy, sensors, excitations of nano-particles, etc. since their technological manufacturing is already available.

%%%%%%%%%%%%%%%%%%%%%%%%%%%%%%%%%%%%%%%%%%%%%%%%%%%%%%%%%%%%%%
\end{document}